\documentclass[aip,apl,reprint,superscriptaddress,floatfix,amsmath,showpacs]{revtex4-1}

\pdfoutput=1

\usepackage{graphicx}
\usepackage{upgreek}
\usepackage{amssymb}
\usepackage{textcomp}
\usepackage{amsmath}
\usepackage{siunitx}

\newcommand{\mum}{\ensuremath{\upmu\text{m}}}
\newcommand{\CMFS}{\ensuremath{\text{Co}_2\text{Mn}_{0.6}\text{Fe}_{0.4}\text{Si}}}

\newcommand{\STTE}{Spin-Transfer-Torque effect}
\newcommand{\SHE}{Spin-Hall-Effect}
\newcommand{\layers}{\mbox{Cr (\SI{5}{nm})$\vert$CMFS (\SI{5}{nm})$\vert$Pt (\SI{2}{nm})}}

\newcommand{\Ith}{\ensuremath{j_\text{Th}}}
\newcommand{\uBLS}{\ensuremath{\upmu}BLS}

\newcommand{\jmmm}{J.\ Magn.\ Magn.\ Mater.\ }

\begin{document}

\title{Characterization of \STTE\ induced magnetization dynamics driven by short current pulses}

\author{T. Meyer}
\affiliation{Fachbereich Physik and Landesforschungszentrum OPTIMAS, Technische Universit\"at
Kaiserslautern, 67663 Kaiserslautern, Germany}
\author{T. Br\"acher}
\affiliation{Fachbereich Physik and Landesforschungszentrum OPTIMAS, Technische Universit\"at
Kaiserslautern, 67663 Kaiserslautern, Germany}
\author{F. Heussner}
\affiliation{Fachbereich Physik and Landesforschungszentrum OPTIMAS, Technische Universit\"at
Kaiserslautern, 67663 Kaiserslautern, Germany}
\author{A.A. Serga}
\affiliation{Fachbereich Physik and Landesforschungszentrum OPTIMAS, Technische Universit\"at
Kaiserslautern, 67663 Kaiserslautern, Germany}
\author{H. Naganuma}
\affiliation{Department of Applied Physics, Graduate School of Engineering, Tohoku University, Sendai 980-8579, Japan}
\author{K. Mukaiyama}
\affiliation{Department of Applied Physics, Graduate School of Engineering, Tohoku University, Sendai 980-8579, Japan}
\author{M. Oogane}
\affiliation{Department of Applied Physics, Graduate School of Engineering, Tohoku University, Sendai 980-8579, Japan}
\author{Y. Ando}
\affiliation{Department of Applied Physics, Graduate School of Engineering, Tohoku University, Sendai 980-8579, Japan}
\author{B. Hillebrands}
\affiliation{Fachbereich Physik and Landesforschungszentrum OPTIMAS, Technische Universit\"at
Kaiserslautern, 67663 Kaiserslautern, Germany}
\author{P. Pirro}
\affiliation{Fachbereich Physik and Landesforschungszentrum OPTIMAS, Technische Universit\"at
Kaiserslautern, 67663 Kaiserslautern, Germany}

\date{\today}

\begin{abstract}
We present a time-resolved study of the magnetization dynamics in a microstructured Cr$\vert$Heusler$\vert$Pt waveguide driven by the \SHE\ and the \STTE\ via short current pulses. In particular, we focus on the determination of the threshold current at which the spin-wave damping is compensated. We have developed a novel method based on the temporal evolution of the magnon density at the beginning of an applied current pulse at which the magnon density deviates from the thermal level. Since this method does not depend on the signal-to-noise ratio, it allows for a robust and reliable determination of the threshold current which is important for the characterization of any future application based on the \STTE.
\end{abstract}


\maketitle
In the last years, the field of magnon spintronics~\cite{Chumak2015} has attracted prominent interest since it offers great potential to realize future logic devices. In this field, currents of magnons, the quanta of spin-waves, are used to transport and process information. Recently, a set of prototype devices has been realized such as, e.g., the magnonic majority gate~\cite{Fischer2017,Klingler2014}, in which the logic state of the output signal is determined by the majority of the input states. Other works on magnon based devices such as the magnon transistor~\cite{Chumak2014}, the spin-wave multiplexer~\cite{Vogt2014}, domain walls as spin-wave nanochannels~\cite{Wagner2016}, graded-index magnonics~\cite{Davies2015} or a signal splitter based on caustic-like spin-wave beams~\cite{Heussner2017} pave the way towards a controlled spin-wave transfer in magnonic networks.

However, the application of these devices is limited by the finite magnon lifetime and, hence, the limited magnon propagation length determined by the Gilbert damping~\cite{Gilbert2004}. Thus, an efficient amplification of spin-waves or a control of the effective damping is inevitable. The latter can be achieved by employing the \STTE~(STT)~\cite{Slonczewski1996}. This effect yields an additional torque on the magnetization which can be co-aligned with the Gilbert damping torque resulting in an effective spin-wave damping given by the interplay of the Gilbert damping and the STT effect. This led, e.g., to the development of spin-torque nano-oscillators based on point-contacts which allow to generate oscillations in the GHz-range by DC currents~\cite{Silva2008,Kim2012}. Furthermore, in combination with the \SHE~(SHE)~\cite{Hirsch1999} to convert a charge current into a spin current in a normal metal, the SHE-STT effect allows for the control of the effective spin-wave damping also in spatially extended magnonic devices~\cite{Ando2008,Demidov2012,Evelt2016,Demidov2011,Gladii2016,An2014}. To achieve large SHE-STT induced effects with minimal current densities, the use of new materials with a low spin-wave damping, such as cobalt-based Heusler compounds~\cite{Mizukami2015}, is very promising. Using the SHE, e.g. in a Pt layer adjacent to a cobalt-based Heusler compound, the SHE-STT effect offers a powerful link to combine magnonics with CMOS technologies.

Up to now, most of the reported studies on the manipulation of spin-waves in a magnonic waveguide via the SHE-STT effect only use static DC currents or quasi-DC pulses with pulse durations of several microseconds. On this time scale, the spin-wave dynamics already reach a quasi-equilibrium state. In contrast, in this Letter, the focus of interest is on the dynamics of the non-equilibrium state during short applied current pulses since the spin-wave properties can be expected to differ strongly from the properties in the quasi-equilibrium state. Furthermore, any future application based on magnons needs to operate on short time scales. Thus, the investigation of the temporal evolution of the spin-wave intensity under the influence of the SHE-STT effect is fundamental and can provide a more detailed insight in the underlying physics and limitations. Hence, the main focus of the presented work is on the temporal dependence of the spin-wave response on \SI{50}{ns} long current pulses which results in a modulation of the effective spin-wave damping via the SHE-STT effect in microstructured Heusler spin-wave waveguides.

In this Letter, we show that in the case of short applied current pulses, the threshold current~\Ith, at which the spin-wave damping is compensated, can be determined via the temporal evolution of the spin-wave intensity during the current pulse once the spin-wave damping is overcompensated. We show that this method is more reliable and robust than the commonly used method for the determination of \Ith\ which is based on the current-dependent inverse spin-wave intensity for currents far below the threshold current~\cite{Demidov2011,Slavin2009}. This conventional method requires a linear equilibrium state of the magnonic system, hence, without any nonlinear spin-wave dynamics. However, assuming an experimental technique with a low temporal resolution and short applied current pulses or a method which delivers a time-integrated spin-wave intensity during an applied current pulse, a nonlinear dynamical equilibrium state during a current pulse can be easily misinterpreted as a linear state which can lead to a strong overestimation of \Ith. In addition, in the case of a small signal-to-noise ratio, the noise, e.g. caused by the dark count rate of the detector, needs to be carefully taken into account when using the conventional method since it strongly influences the inverse spin-wave intensity and, hence, affects the obtained values for \Ith. In contrast, the developed method employing the temporal evolution of the spin-wave intensity is, in general, independent on the noise level. Furthermore, since the developed method requires an overcompensation of the spin-wave damping, the spin-wave intensity is drastically increased compared to the intensity in the absence of a current pulse which improves the signal-to-noise ratio. This, in return, might also allow for the determination of \Ith\ via the developed method, e.g., using the inverse \SHE~\cite{Chumak2015}.

\begin{figure}[t]
\includegraphics[width=0.48 \textwidth ]{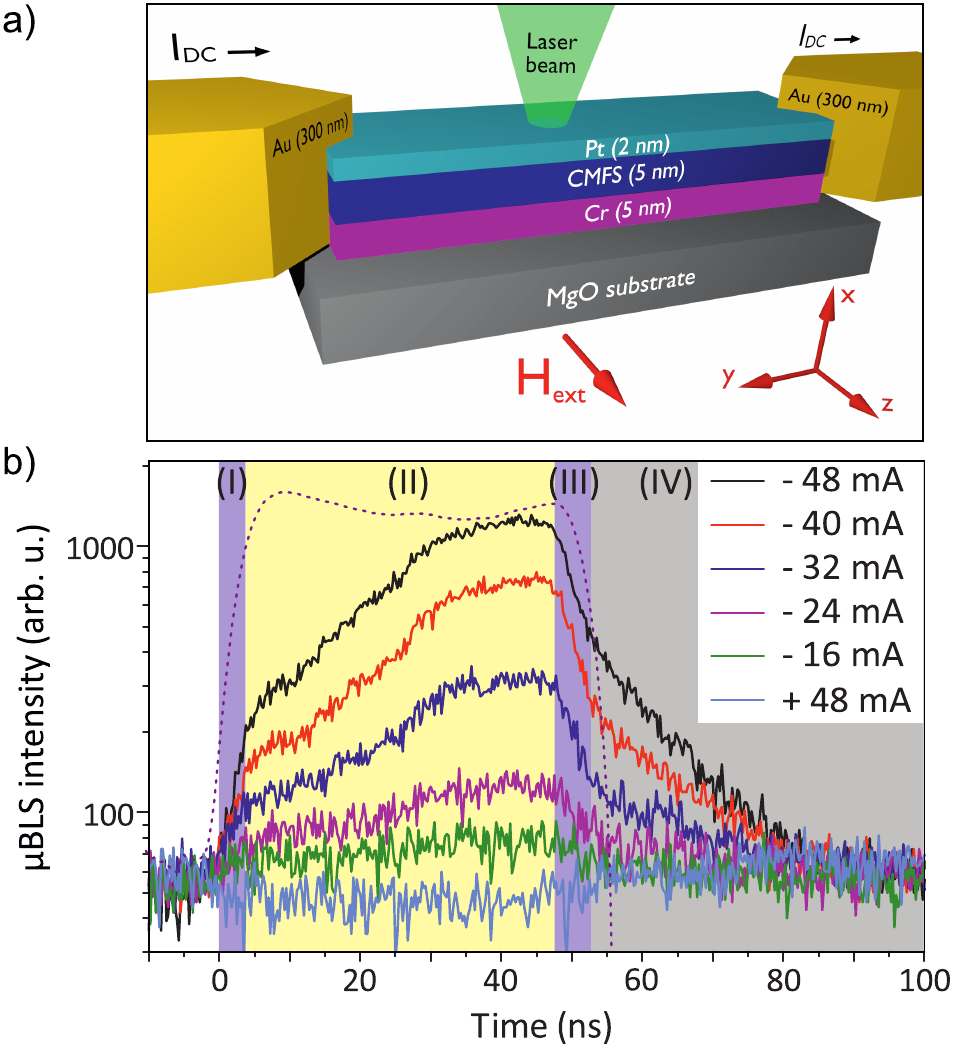}
\caption{\label{Fig:sample}(color online) a) Sketch of the sample and the used coordinate system. The \SI{7}{\mum} wide and \SI{30}{\mum} long waveguide consists of a \layers\ trilayer, and Au contacts at each end allow for the application of a charge current to the layer stack. The magnetization is aligned along the z-direction. b) Time evolution of the \uBLS\ intensity for \SI{50}{ns} long applied current pulses and different current values (note the log-scale). The dashed line indicates the shape of the current pulse and the shaded areas indicate the different regimes of the temporal evolution of the spin-wave intensity.}
\end{figure}

In the following, time-resolved Brillouin light scattering microscopy~(\uBLS) is employed to observe the spin-wave dynamics in a spin-wave waveguide~\cite{Sebastian2015}. The magnetic layer of the \SI{7}{\mum} wide and \SI{30}{\mum} long waveguide is made from the low-damping Heusler compound \CMFS~(CMFS)~\cite{Sebastian2012} (see Fig.~\ref{Fig:sample}a). To support the crystalline growth of the CMFS layer, a \SI{5}{nm} thick Cr layer is deposited on a MgO substrate, acting as a buffer layer. On top, a \SI{5}{nm} thick CMFS layer is grown. Finally, a Pt layer of \SI{2}{nm} thickness is deposited. The subsequent microstructuring into waveguides is performed by means of electron beam lithography and ion beam etching.

In order to apply charge currents, \SI{300}{nm} thick Au contacts are patterned at each end of the waveguide. The applied current flows through the trilayer and, via the SHE, generates a spin current flowing perpendicular to the charge current. By aligning the magnetization of the CMFS layer via an external magnetic field as shown in Fig.~\ref{Fig:sample}a, the injected spin current produces a torque on the precessing CMFS moments via the SHE-STT effect which is co-aligned with the Gilbert damping torque~\cite{Krivorotov2005}.

Since all layers of the waveguide are metallic, the applied charge current will be distributed among them. Thus, not only the charge current in the Pt layer but also the current in the Cr layer will be partially converted into a pure spin current, and, subsequently, both spin currents are injected into the CMFS layer. According to Ref. \onlinecite{Du2014}, Cr exhibits a negative spin-Hall angle (SHA)~\cite{Schreier2014}, whereas the SHA of Pt is positive. Due to this fact and since the spin currents emerging from the Pt and Cr layers enter the CMFS layer from opposite surfaces, both spin currents add up. In contrast, the Oersted fields induced in the CMFS layer by the electric currents flowing in the Pt and Cr layers have different signs. Since they are similar in magnitude, they do not influence the spin-wave properties significantly.

Depending on the current direction, the reduced (increased) damping leads to an increase (decrease) from the thermal equilibrium magnon density which can be observed by means of \uBLS\ spectroscopy of the thermal magnons. For the following measurements, the probing laser spot with a diameter of approximately \SI{400}{nm} is positioned in the center of the waveguide (see Fig.~\ref{Fig:sample}a).

Figure~\ref{Fig:sample}b shows the time-resolved \uBLS\ intensity measured for an external magnetic field of \mbox{${\upmu_0}{H_\text{ext}}=+70\text{~mT}$} for \SI{50}{ns} long applied current pulses with different current values~\cite{current}. Exemplarily, the shape of the current pulse is shown by the dashed line. In general, four distinct regimes of the induced spin-wave dynamics can be identified as indicated by the colored areas. Regimes~(I) and (III) relate to the first and last part ($\approx$ \SI{5}{ns}) of the current pulse, respectively. Here, the \uBLS\ intensity rapidly changes, indicating the transition to a new state. The rise (fall) time of this transition is in the same order as the rise (fall) time of the current pulse, respectively. Thus, this transition happens on a faster time scale as the time-resolution of the experimental setup which is in the order of~\SI{1}{ns}. Regime~(II) corresponds to the time interval during which the applied current is approximately constant. Thus, also the torque acting on the magnetization is constant~\cite{torque-temperature,Uchida2014}. Finally, regime~(IV) describes the evolution of the \uBLS\ intensity after the current pulse.

As can easily be seen from Fig.~\ref{Fig:sample}b, the intensity increases in the case of negative applied currents which refers to a reduction of the effective spin-wave damping. In contrast, for positive applied currents, thermally excited spin-waves can be suppressed~\cite{Demidov2011,Meyer2017} as exemplarily shown by the reduced intensity during the current pulse for an applied current of \mbox{$j=+48\text{~mA}$} (light blue curve). 

Based on this data, the threshold current~\Ith\ is determined which is one of the most important parameters for any system driven by the SHE-STT effect. As already mentioned, for this, a novel method is presented which is based on the temporal evolution of the spin-wave intensity which is in analogy to a parametric amplification of spin-waves~\cite{Brächer2017,Ulrichs2011}. Here, the growth rate of the spin-wave intensity in the beginning of regime~(II) is taken into account. In the case of an overcompensated effective spin-wave damping by the SHE-STT effect, the spin-wave intensity is expected to drastically increase compared to a reduced but still positive effective spin-wave damping. Exemplarily, this can be seen from Fig.~\ref{Fig:sample}b (note the log-scale) since for large negative currents an exponential increase of the \uBLS\ intensity is obtained which is caused by a negative effective spin-wave damping.

In the following, the spin-wave damping is represented by the spin-wave relaxation rate~$\Gamma$ to allow for a simple description of the temporal evolution of the spin-wave intensity. Furthermore, the SHE-STT effect influences the effective spin-wave damping which is taken into account by the current-dependent parameter~$\beta(j)$. This yields an effective magnon relaxation rate given by $(\Gamma-\beta(j))$ in regime~(II) since the torque acting on the magnetization via the SHE-STT torque is constant during this time.

Considering a negative effective spin-wave damping, hence, $\Gamma < \beta(j)$, this leads to an exponential increase of the \uBLS\ intensity $I_{\upmu \text{BLS}}(t)$. Thus, in the beginning of regime(II), hence, in the case of an increase of the magnon density starting from the thermal level and before any saturation effects occur, $I_{\upmu \text{BLS}}(t)$ is given via:
\begin{equation}
{{I_{\upmu \text{BLS}}}(t) \propto \text{e}^{g(j) t}}.
\label{Eq:Risetime}
\end{equation}
Here, $g(j>\Ith) = \beta(j)-\Gamma > 0$ denotes the growth rate of the spin-wave intensity. In contrast, in the case of a positive effective spin-wave damping with \mbox{$\beta(j)-\Gamma < 0$}, Eq.~\ref{Eq:Risetime} is not valid any more and $I_{\upmu \text{BLS}}(t)$ should be constant in regime~(II) resulting in a vanishing \mbox{$g(j<\Ith) \approx 0$}.

For the sake of completeness, it should be noted that also for currents below \Ith, the spin-wave intensity shows a slight increase over time during the pulse duration. This is caused, e.g., due to an increasing magnon density caused by an increased temperature due to Joule heating~\cite{Meyer2017}. However, in this case, $I_{\upmu \text{BLS}}$ does not diverge in the beginning of the pulse but always show a saturation-like behavior.
\begin{figure}[t]
\includegraphics[width=0.48 \textwidth ]{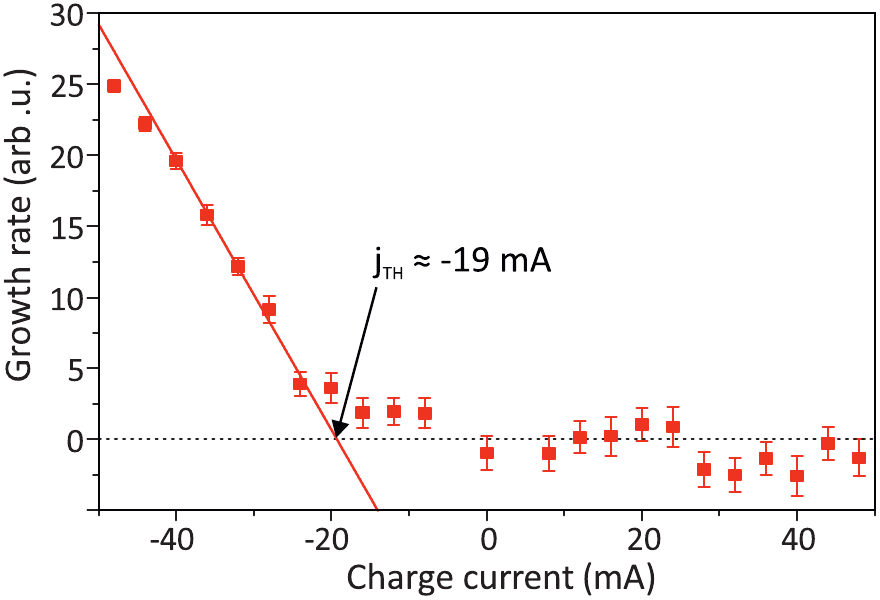}
\caption{\label{Fig:GrowthRate}(color online) Current dependent growth rate of the \uBLS\ intensity determined via exponential fits to the corresponding temporal evolution in regime~(II). The solid line depicts a linear fit to the data for currents below \SI{-20}{mA}.}
\end{figure}

For the subsequent determination of \Ith, exponential fits are applied to the spin-wave intensity in the beginning of regime~(II). In combination with the previous considerations, the so obtained exponential growth rates should only yield a non-zero value in the case of a negative effective spin-wave damping with $g(j>\Ith) > 0$. The results are shown in Fig.~\ref{Fig:GrowthRate} and, indeed, for positive applied currents and small negative applied currents, the obtained growth rates vanish while for large negative applied currents, \Ith\ is overcome leading to an exponential increase of the spin-wave intensity during the current pulse.

Furthermore, $g(j)$ should scale linearly with $j$ since $\beta(j) \propto j$~\cite{Demidov2011,Slavin2009}. Thus, \Ith\ can be easily obtained by applying a linear fit to the obtained current dependent growth rates for large negative applied currents (solid line in Fig.~\ref{Fig:GrowthRate}) yielding $\Ith \approx -19~\text{mA}$.

Thus, in conclusion, the presented method for the determination of \Ith\ allows for a reliable determination of \Ith\ in the case of a system driven by the SHE-STT effect using short current pulses. 

In contrast, as mentioned before, assuming a measurement technique with a lower time-resolution, the spin-wave intensity level reached at the end of the pulse in Fig.~\ref{Fig:sample}b, might be easily misinterpreted as the reaching of an equilibrium state with a reduced but still positive effective spin-wave damping. This might lead to the false assumption that the requirements of a linear equilibrium state with a positive effective spin-wave damping for the determination of \Ith\ via the commonly used method are fulfilled. In contrast, from the previously presented results, it is obvious that for large negative currents, the saturation-level of the spin-wave intensity is determined by nonlinear effects, thus, the conventional method is not valid in this current regime. However, as demonstrated in the following, even though the application of this method experimental data for currents above \Ith\ is not valid any more, the obtained results show the expected behavior which is misleading and yields a strongly overestimated value for \Ith.

\begin{figure}[t]
\includegraphics[width=0.48 \textwidth ]{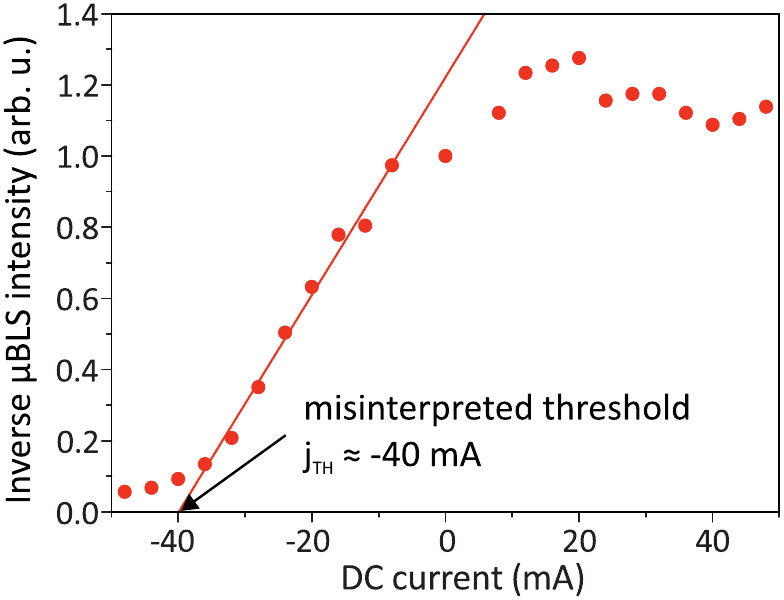}
\caption{\label{Fig:Data}(color online) Normalized inverse \uBLS\ intensity (red circles). The solid line represents a linear fit to the data points in the range of \SI{-8}{mA} to \SI{-36}{mA}.}
\end{figure}

For the conventional determination of \Ith, the energy in the spin-wave system is considered. According to theory~\cite{Demidov2011,Slavin2009}, for a positive effective spin-wave damping and for a small spin-wave amplitude, hence, as long as no nonlinear damping processes occur, the inverse equilibrium spin-wave intensity scales linearly with the effective spin-wave damping:
\begin{equation}
I_{\upmu \text{BLS}}^{-1}(j) \propto \Gamma-\beta(j).
\label{Eq:Threshold}
\end{equation}
Similar to the presented determination via the temporal evolution of the spin-wave intensity, this yields a linear dependence of $I_\text{\uBLS}^{-1}(j)$ on $j$ for small spin-wave intensities~\cite{Demidov2011}. Based on this, \Ith\ is determined via a linear extrapolation of $I_\text{\uBLS}^{-1}(j)$ as the current at which \mbox{$I_{\upmu \text{BLS}}^{-1}(\text{j}_\text{Th})=0$} as indicated in Fig.~\ref{Fig:Data}.

It should be noted that Eq.~\ref{Eq:Threshold} can only be applied if the noise level is negligible. However, for small negative applied currents and for positive applied charge currents, the noise level might influence the inverse \uBLS\ intensity. In particular, the decrease of the inverse \uBLS\ intensity for currents above \SI{+20}{mA} is mainly caused due to an enhanced thermal magnon density caused by an increased sample temperature due to Joule heating as demonstrated in~\cite{Meyer2017}. Since this influence of the thermal magnon density is not caused via the SHE-STT effect, this influence needs also to be taken into account as an additional current-dependent noise level.

As mentioned before, assuming an experimental method with a lower time-resolution, the reached saturation level during a short current pulse can be easily misinterpreted since a linear or a nonlinear saturation cannot be easily distinguished from the experimental data. In particular, the rather linear dependence of the inverse \uBLS\ intensity on $j$ for $-36~\text{mA}<j<-8~\text{mA}$ as depicted in Fig.~\ref{Fig:Data}, might indicate that Eq.~\ref{Eq:Threshold} is still valid in this current range, leading to a strongly overestimated and misinterpreted threshold current of approximately \SI{-40}{mA}.

To understand the observed linear dependence of the inverse \uBLS\ intensity on $j$ even though Eq.~\ref{Eq:Threshold} is not valid in this regime, one needs to consider that the injected spin current via the SHE is finite. Thus, the total angular momentum injected into the magnonic system is also finite which results in a limited spin-wave intensity which can be sustained even though the effective spin-wave damping is negative. In addition, the injected spin current scales with $j$, which, in return, yields the observed linear dependence of the inverse \uBLS\ intensity in Fig.~\ref{Fig:Data}. Thus, both mechanisms, a decrease of the (still positive) effective spin-wave damping as well as the finite injected spin current, lead to a similar dependence of the inverse \uBLS\ intensity on $j$. 

In summary, in this Letter, a novel method for the determination of the threshold current for SHE-STT driven systems via short current pulses was presented. By evaluating the temporal evolution of the spin-wave intensity in the case of an overcompensated effective spin-wave damping, the developed method yields a reliable determination of \Ith. In contrast, it could be demonstrated that for the commonly used method for the determination of \Ith\ based on a linear extrapolation of the inverse spin-wave intensity, other effects need to be carefully taken into account which affect the determination. In particular, the consequences of a misinterpretation of the occurrence of a saturation leads to a strong overestimation of \Ith\ via the conventional method. In return, our novel method does not require an equilibrium state and is less affected by other influences like, e.g., the signal-to-noise ratio and, hence, constitutes a more robust and reliable method for the determination of \Ith. Thus, the presented results show that a time-resolved detection technique is inevitable for a reliable characterization of a system driven by the SHE-STT effect via short current pulses.

\begin{acknowledgments}
The authors gratefully acknowledge financial support by the DFG in the framework of the research unit TRR 173 ``Spin+X'' and by the DFG Research Unit 1464 and the Strategic Japanese-German Joint Research Program from JST: ASPIMATT.
\end{acknowledgments}


\begin{thebibliography}{}
\bibitem{current}It should be noted that in the experiment, a voltage pulse is applied to the waveguide and the current is derived by the DC resistance of the device of $\approx$ \SI{250}{\Omega} using Ohm's law. Due to the impedance mismatch the actual current might slightly deviate from the indicated currents by a scaling factor.
\bibitem{torque-temperature}In general, the spin-mixing conductance, and, consequently, the injected spin current and the torque on the magnetization are temperature dependent. However, in the following, the temperature dependence of the torque is neglected since all observed phenomena can be explained by the increased magnon density.
\end{thebibliography}
\end{document}